\documentclass[twocolumn,showpacs]{revtex4}
\usepackage{amsfonts,amsmath,amssymb}
\usepackage{color}
\usepackage[dvips]{graphicx}
\usepackage{dcolumn}
\usepackage{bm}
\usepackage{lscape}

\begin{document}

\title{Mott-Hubbard Transition of Bosons in Optical Lattices with Three-body
Interactions}
\author{Bo-Lun Chen}
\email{bolun_chen@mail.bnu.edu.cn}
\affiliation{Department of Physics, Beijing Normal University, Beijing 100875, China}
\author{Xiao-Bin Huang}
\affiliation{Department of Physics, Beijing Normal University, Beijing 100875, China}
\author{Su-Peng Kou}
\thanks{Corresponding author}
\email{spkou@bnu.edu.cn}
\affiliation{Department of Physics, Beijing Normal University, Beijing 100875, China}
\author{Yunbo Zhang}
\affiliation{Department of Physics and Institute of Theoretical Physics, Shanxi
University, Taiyuan 030006, China}

\begin{abstract}
In this paper, the quantum phase transition between superfluid state and
Mott-insulator state is studied based on an extended Bose-Hubbard model with
two- and three-body on-site interactions. By employing the mean-field
approximation we find the extension of the insulating 'lobes' and the
existence of a fixed point in three dimensional phase space. We investigate
the link between experimental parameters and theoretical variables. The
possibility to obverse our results through some experimental effects in
optically trapped Bose-Einstein Condensates(BEC) is also discussed.
\end{abstract}

\maketitle

\section{INTRODUCTION}

Quantum phase transitions in ultracold atoms(BEC) exhibit fascinating
phenomena and attract much attention since such a system with controllable
interactions between atoms had been realized experimentally by loading the
condensate\ into an optical lattice \cite{M. Greiner}. The behavior of the
BEC in optical lattices can be well described by the Bose-Hubbard model \cite%
{Fisher,Scalettar,Niyaz,Rokhsar,Krauth,Batrouni,Sheshadri,Bruder,Freericks,Jaskch,kou}
which predicts a superfluid-insulator transition under certain critical
conditions for the interatomic interaction strengths. The rich phenomena as
well as sophisticated phase diagrams resulting from quantum many-body
effects have been explored plentifully \cite{Amico,Elstner,Sachdev,D.van
Oosten,W.Zwerger,Goral,Liang,W.Zwerger's Review}. Most previous discussions
assumed that interactions between three and(or) more trapped bosons are
negligible so the system is merely controlled by the two-body interaction
and tunneling effect which allow bosons to hop around within the optical
lattice.

However, it has been an intriguing question of pursuing some exotic phases
associated with Hamiltonians with three- or more-body terms \cite{T. Kohler}
for a long time. Quite recently, B\"{u}chler \textit{et al.} \cite{H. P.
Buchler} suggested that polar molecules in optical lattices driven by
microwave fields naturally give rise to Hubbard model with strong
nearest-neighbor three-body interaction, while the two-body term can be
tuned with external fields. This may lead to some interesting quantum phases
when generalized to more complicated situations.

In the present work, we discuss the superfluid to Mott-insulator transition
of cold bosons in optical lattices with two-and three-body repulsive
interactions. We find that in the strong coupling limit, an effective
on-site Hamiltonian can be derived with a decoupling\ mean-field
approximation developed in Ref. \cite{D.van Oosten}. We then calculate the
energy of this Hamiltonian using perturbation theory and discuss the
critical behavior at transition point\ with Landau order parameter
expansion. We find that due to the presence of the three-body interaction,
the Mott-insulator areas expand to three-dimensional curved surfaces, and
these 3D insulating 'lobes' rotate with respect to a fixed point in the
phase space, as more atoms are populated on the lattice sites. We then
analyze some experimental parameters corresponding to the theoretical
quantities discussed here and suggest some practical ways to verify our
results.

\section{MEAN-FIELD APPROXIMATION}

In this section, we first derive the effective Hamiltonian in the mean-field
approximation and then calculate the energy of this Hamiltonian by applying
perturbation theory up to the fourth order. The equation of critical curves
separating the Mott insulator and superfluid phases are obtained through
Landau order parameter expansion for second-order phase transitions.

\subsection{The Effective Hamiltonian}

We assume that the atoms are in the lowest band of the optical lattice, thus
their behaviors can be described by an (extended) Bose-Hubbard Hamiltonian
\begin{eqnarray}
\hat{H} &=&-t\underset{\left\langle i,j\right\rangle }{\sum }\hat{c}%
_{i}^{\dagger }\hat{c}_{j}+\frac{U}{2}\underset{i}{\sum }\hat{n}_{i}(\hat{n}%
_{i}-1)  \notag \\
&&+\frac{W}{6}\underset{i}{\sum }\hat{n}_{i}(\hat{n}_{i}-1)(\hat{n}%
_{i}-2)-\mu \underset{i}{\sum }\hat{n}_{i}\text{,}  \label{H}
\end{eqnarray}%
where the summation in the first term is done for the nearest neighbors only
and $\hat{c}_{i}^{\dagger }(\hat{c}_{i})$ is the bosonic
creation(annihilation) operator with $\hat{n}_{i}$ the particle number
operator for bosons at site $i$. The parameter $t$ is the hopping
term(tunneling amplitude) and $U(W)$ is two(three)-body repulsive
interaction strength between bosons. The chemical potential $\mu $ is
introduced to conserve the number of atoms in the grand-canonical ensemble.

In the \emph{strong coupling limit}, namely, $t\ll \min \left\{ U,W\right\} $%
, it is convenient to introduce a superfluid order parameter \cite{D.van
Oosten} $\psi =\sqrt{\langle \hat{n}_{i}\rangle }=\langle \hat{c}%
_{i}^{\dagger }\rangle =\langle \hat{c}_{i}\rangle $ to construct a
consistent mean-field approximation by substituting the hopping term in Eq. (%
\ref{H}) as
\begin{equation}
\hat{c}_{i}^{\dagger }\hat{c}_{j}=\langle \hat{c}_{i}^{\dagger }\rangle \hat{%
c}_{j}+\hat{c}_{i}^{\dagger }\langle \hat{c}_{j}\rangle -\langle \hat{c}%
_{i}^{\dagger }\rangle \langle \hat{c}_{j}\rangle =\psi (\hat{c}%
_{i}^{\dagger }+\hat{c}_{j})-\psi ^{2}\text{.}  \label{eqn2}
\end{equation}%
Thus the effective Hamiltonian becomes
\begin{eqnarray}
\hat{H}^{\text{eff}} &=&-zt\psi \underset{i}{\sum }(\hat{c}_{i}^{\dagger }+%
\hat{c}_{i})+zt\psi ^{2}N_{s}+\frac{U}{2}\underset{i}{\sum }\hat{n}_{i}(\hat{%
n}_{i}-1)  \notag \\
&&+\frac{W}{6}\underset{i}{\sum }\hat{n}_{i}(\hat{n}_{i}-1)(\hat{n}%
_{i}-2)-\mu \underset{i}{\sum }\hat{n}_{i}\text{,}  \label{Heff}
\end{eqnarray}%
where $z$ is the coordination number of optical lattices and $N_{s}$ is the
total number of lattice sites. It can be diagonalized and rewritten with
respect to the site index $i$ as
\begin{eqnarray}
\hat{H}_{i}^{\text{eff}} &=&\frac{\overline{U}}{2}\hat{n}_{i}(\hat{n}_{i}-1)+%
\frac{\overline{W}}{6}\hat{n}_{i}(\hat{n}_{i}-1)(\hat{n}_{i}-2)  \notag \\
&&-\overline{\mu }\hat{n}_{i}-\psi (\hat{c}_{i}^{\dagger }+\hat{c}_{i})+\psi
^{2}\text{,}  \label{Heffi}
\end{eqnarray}%
where dimensionless parameters $\overline{U}=U/zt$, $\overline{W}=W/zt$ and $%
\overline{\mu }=\mu /zt$ illustrate relative strengths between repulsive
interactions (two- and three-body) and tunneling effect (the hopping term $t$%
). This on-site Hamiltonian can be applied equally to every lattice site so
we will omit the index $i$ in following discussions.

\subsection{Perturbation Theory}

In the strong coupling regime, we apply perturbation theory to calculate the
energy of Eq. (\ref{Heffi}). To achieve this, we rewrite Eq. (\ref{Heffi})\
as
\begin{equation}
\hat{H}^{\text{eff}}=\hat{H}^{\left( 0\right) }+\psi \hat{V}  \label{eqn5}
\end{equation}%
with the unperturbed\ Hamiltonian $\hat{H}^{\left( 0\right) }=\frac{%
\overline{U}}{2}\hat{n}(\hat{n}-1)+\frac{\overline{W}}{6}\hat{n}(\hat{n}-1)(%
\hat{n}-2)-\overline{\mu }\hat{n}+\psi ^{2}$, and the perturbation $\hat{V}%
=-(\hat{c}_{i}^{\dagger }+\hat{c}_{i})$, representing the small kinetic
energy of bosons near the insulating phase. The ground-state energy of $\hat{%
H}^{\left( 0\right) }$ is
\begin{eqnarray}
E_{g}^{(0)} &=&\left\{ E_{n}^{(0)}|_{n=0,1,2,\ldots }\right\} _{\min }
\notag \\
&=&\frac{\overline{U}}{2}g(g-1)+\frac{\overline{W}}{6}g(g-1)(g-2)-\overline{%
\mu }g\text{,}  \label{Eg}
\end{eqnarray}%
where $E_{n}^{(0)}$ denotes the unperturbed energy of states with integer
fillings and $g$ is an integer specifying the average particle number on a
lattice site.

In the occupation number basis, the odd powers of the expansion of Eq. (\ref%
{eqn5}) with respect to $\psi $ vanish. Therefore, the correction to the
energy can be calculated to the second-order
\begin{equation}
E_{g}^{(2)}=\psi ^{2}\underset{n\neq g}{\sum }\frac{\left\vert \left\langle
g\left\vert \hat{V}\right\vert n\right\rangle \right\vert ^{2}}{%
E_{g}^{(0)}-E_{n}^{(0)}}=a_{2}\psi ^{2}  \label{Eg2}
\end{equation}%
with
\begin{eqnarray}
a_{2} &=&\left[ \frac{g}{\left( g-1\right) \overline{U}+\frac{1}{2}\left(
g-1\right) \left( g-2\right) \overline{W}-\overline{\mu }}\right.   \notag \\
&&\left. +\frac{g+1}{\overline{\mu }-g\overline{U}-\frac{1}{2}g\left(
g-1\right) \overline{W}}\right] ,
\end{eqnarray}%
and to the fourth-order
\begin{align}
E_{g}^{(4)}& =\underset{n,p,q\neq g}{\sum }\left\vert \left\langle
g\left\vert \hat{V}\right\vert n\right\rangle \right\vert \left[ -E_{g}^{(2)}%
\frac{\left\vert \left\langle n\left\vert \hat{V}\right\vert g\right\rangle
\right\vert }{\left( E_{g}^{(0)}-E_{n}^{(0)}\right) ^{2}}\right.   \notag \\
& \left. +\frac{\left\vert \left\langle n\left\vert \hat{V}\right\vert
p\right\rangle \right\vert }{E_{g}^{(0)}-E_{n}^{(0)}}\frac{\left\vert
\left\langle p\left\vert \hat{V}\right\vert q\right\rangle \right\vert }{%
E_{g}^{(0)}-E_{p}^{(0)}}\frac{\left\vert \left\langle q\left\vert \hat{V}%
\right\vert g\right\rangle \right\vert }{E_{g}^{(0)}-E_{q}^{(0)}}\right]
\notag \\
& =a_{4}\psi ^{4}  \label{Eg4}
\end{align}%
with
\begin{widetext}
\begin{align}
a_{4}& =\frac{g(g-1)}{\left[ \left( g-1\right) \overline{U}+\frac{1}{2}%
\left( g-1\right) \left( g-2\right) \overline{W}-\overline{\mu }\right] ^{2}%
\left[ (2g-3)\overline{U}+(g-2)^{2}\overline{W}-2\overline{\mu
}\right] }
\notag \\
& \text{ \ \ \ }+\frac{\left( g+1\right) \left( g+2\right) }{\left[
\overline{\mu }-g\overline{U}-\frac{1}{2}g\left( g-1\right) \overline{W}%
\right] ^{2}\left[ 2\overline{\mu }-\left( 2g+1\right) \overline{U}-g^{2}%
\overline{W}\right] }  \notag \\
& \text{ \ \ \ }-\left[ \frac{g}{\left( g-1\right) \overline{U}+\frac{1}{2}%
\left( g-1\right) \left( g-2\right) \overline{W}-\overline{\mu }}+\frac{g+1}{%
\overline{\mu }-g\overline{U}-\frac{1}{2}g\left( g-1\right) \overline{W}}%
\right] \times   \notag \\
& \text{ \ \ \ \ \ \ }\left[ \frac{g}{\left( \left( g-1\right) \overline{U}+%
\frac{1}{2}\left( g-1\right) \left( g-2\right) \overline{W}-\overline{\mu }%
\right) ^{2}}+\frac{g+1}{\left( \overline{\mu }-g\overline{U}-\frac{1}{2}%
g\left( g-1\right) \overline{W}\right) ^{2}}\right] \text{.}
\label{a4}
\end{align}%
\end{widetext}Here, $|n\rangle ,|p\rangle ,|q\rangle $ denote unperturbed
states with $n,p,q$ particles.

\subsection{Landau Order Parameter Expansion}

We then follow the ordinary procedure for second order phase transitions by
writing the ground-state energy $E_{g}\left( \psi \right)
=E_{g}^{(0)}+E_{g}^{(2)}+E_{g}^{(4)}+...$ as an expansion in $\psi $, with
an implication that the superfluidity cannot be large in the strong coupling
region:
\begin{eqnarray}
E_{g}\left( \psi \right)  &=&a_{0}\left( g,\overline{U},\overline{W},%
\overline{\mu }\right) +\left[ 1+a_{2}\left( g,\overline{U},\overline{W},%
\overline{\mu }\right) \right] \psi ^{2}  \notag \\
&&+a_{4}\left( g,\overline{U},\overline{W},\overline{\mu }\right) \psi ^{4}+%
\mathcal{O}\left( \psi ^{6}\right) .  \label{eqn9}
\end{eqnarray}

By minimizing Eq. (\ref{eqn9})\ with respect to the superfluid order
parameter $\psi $, we obtain the critical condition $1+a_{2}=0$ signifying
the\ boundary that separates insulator and superfluid phases, which yields
\begin{equation}
\overline{\mu }_{\pm }=\frac{1}{2}\left[ \left( 2g-1\right) \overline{U}%
+\left( g-1\right) ^{2}\overline{W}-1\right] \pm \frac{1}{2}\sqrt{\Delta }
\label{eqn12}
\end{equation}%
with
\begin{eqnarray}
\Delta  &=&\overline{U}^{2}+\left( g-1\right) ^{2}\overline{W}^{2}+2\left(
g-1\right) \overline{U}\overline{W}  \notag \\
&&-2(2g+1)\overline{U}-2(2g+1)(g-1)\overline{W}+1,  \label{delta}
\end{eqnarray}%
where the subscript $\pm $ denotes the upper and lower halves of the Mott
insulating regions in phase space. The parameter $\Delta $ will play a
significant role in determining phase diagrams, which will be discussed in
detail in the\ next section.

\begin{figure*}
\begin{center}
\includegraphics[width=0.95\textwidth]{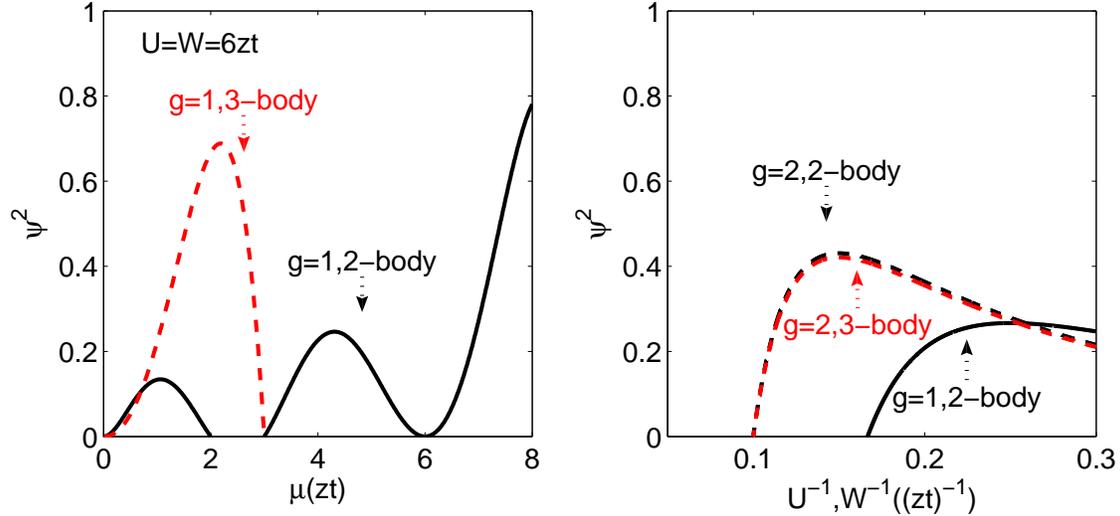}
\caption{(Color online) Illustrations of superfluid order parameters $%
\protect\psi ^{2}$ according to Eq. (\protect\ref{eqn14}). Left: with fixed
repulsion and varying\ chemical potential $\overline{\protect\mu }$ at $g=1$%
. The black line denotes $\overline{U}=6,\overline{W}=0$; The red dash line
denotes $\overline{U}=0,\overline{W}=6$. Right: with $\overline{\protect\mu }%
=(g-\frac{1}{2})\overline{U}+\frac{1}{2}(g-1)^{2}\overline{W}$ and varying $%
\overline{U}^{-1},\overline{W}^{-1}$. The solid line denotes $g=1$,
the dash lines denote $g=2$.} \label{Fig.0}
\end{center}
\end{figure*}

For given particle number $g$, the weak superfluidity is
\begin{equation}
\psi ^{2}=-\frac{1+a_{2}}{2a_{4}}\text{,}  \label{eqn14}
\end{equation}%
which is illustrated in Fig. \ref{Fig.0}. Here, we would like to make a
remark on our decoupling-Landau expansion scheme. This method was originally
introduced\cite{D.van Oosten} as an alternative way of the usual Bogoliubov
approximation, which did not predict the phase transition. Instead of
considering number fluctuations with respect to the number of condensed
atoms in Bogoliubov treatment, decoupling method approximates the kinetic
term as site-independent and weak, thus it reveals the expected transition.
But it only works\ well in the \emph{strong-coupling limit }($U/t,W/t\gg 1$%
), making hopping as perturbations in expansion of $E\left( \psi
\right) $. It cannot describe the system properly in the
weak-coupling limit, as shown in right plot of Fig. \ref{Fig.0}:
$\psi ^{2}$ tends to decrease towards zero as repulsions decline.
Also, in the case of $g=1,U=0$, $\psi ^{2}$ exhibits strange
behavior which is really beyond our approach (the system should be a
weak-interaction BEC in this sense, we shall discuss it elsewhere).
The incorrectness stems from the limited effective range of our
approach.

\section{THE PHASE DIAGRAMS}

In this section, we present the phase diagrams based on results obtained in
the previous section and demonstrate their physical meanings with the help
of Eq. (\ref{delta}).

According to Eq. (\ref{eqn12}), physically acceptable solutions of the
reduced\ chemical potential $\overline{\mu }$ require that $\Delta \geq 0$.
This restriction, which guarantees $\overline{\mu }$ being real-valued and
the existence of Mott insulating phases, can be visualized in Fig. \ref%
{Fig.1} as we set Eq. (\ref{delta}) to be zero. In fact,\ solutions to $%
\Delta =0$ are a set of parallel lines in the $\overline{U}$-$\overline{W}$
plane. They divide the plane into two parts: areas to their right correspond
to $\Delta >0$, which implies the appearance of the Mott-insulating phase;
areas to their left correspond to $\Delta <0$, which implies the appearance
of the superfluid phase. Areas with negative values of ($\overline{U}$ or $%
\overline{W}$) are not shown since we only consider repulsive interactions
among atoms.

\begin{figure}
\begin{center}
\includegraphics[width=0.44\textwidth]{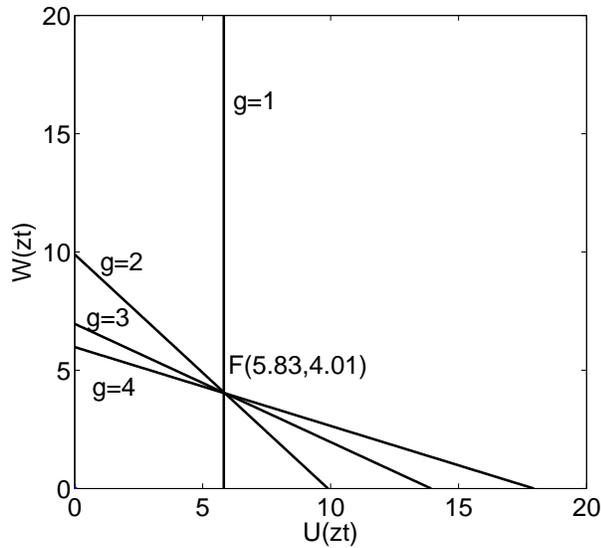}
\caption{Illustration of solutions to $\Delta =0$ with different
occupancy $g$. The critical lines are tilting down as$\ g$
increases\ and all of them intersect around a fixed point point
$F(5.83,4.01) $.} \label{Fig.1}
\end{center}
\end{figure}

The critical line is perpendicular to the $\overline{U}$-axis with an
intercept of $\overline{U}_{C}=5.83$ when $g=1$, which is exactly the
transition point of system with pure\ two-body interaction. It seems to be\
contradictory, since according to Eq. (\ref{H}), both the two-body and the
three-body terms have to\ vanish at $g=1$, and the system ought to\ behave
as a non-interacting BEC. We can understand it in the following way:

Although the\ on-site repulsion $U\hat{n}_{i}(\hat{n}_{i}-1)$ vanishes at
single occupancy, two-body interactions can\ take place due to weak
(perturbative) tunneling effect. It enables one boson to 'jump' from one
site to its nearest neighbors, so the on-site\ repulsion between two bosons
appears again, raising energy and may forbid the occupation of\ an
additional atom if $U$ is strong enough. In short, when hoppings are added
as perturbations (the Landau expansions in $\psi ^{2}$), the condensate can
exhibit a weak\ superfluid phase. In the\ large $U/t$ regime, hoppings are
prohibited and\ the condensate becomes insulating.

Similarly,\ three-body interactions $W\hat{n}_{i}(\hat{n}_{i}-1)(\hat{n}%
_{i}-2)$ may appear as two (or more) bosons successively move onto an
occupied site. But these processes are\ hidden in higher-order terms ($\psi
^{4}$ and above)\ in our kinetic-perturbation expansion, thus can be
neglected here. This explains the absence of three-body interactions at $g=1$%
. Hence,\ the condensate can be described as a weak-interacting BEC with\
finite two-body repulsions and can have the Mott transition.

When $g=2$, the $U$-term is recovered while the $W$-term can\ emerge\ by
adding hopping perturbations (with one atom jumping). So the system can
either be a weak superfluid or an\ insulator, depending on the ratio $U/t$
and $W/t$.

In addition, three-body term will surpass the two-body one gradually as $g$
increases. From Fig. \ref{Fig.1}, we can see that as there are more
particles on sites, intercepts of the critical lines to $\overline{W}$-axis
become smaller; when $g\rightarrow \infty $, the intercept goes to $%
\overline{W}_{C}=4.01$. In fact, all the lines with different $g$ intersect
around the point $F(5.83,4.01)$ and rotate with respect to it: from being
perpendicular to the $\overline{U}$-axis when $g=1$ to being parallel to the
$\overline{U}$-axis when $g\rightarrow \infty $.

This behavior can be straightforwardly explained from the dispersion
relation (obtained in a similar way as in Ref. \cite{D.van Oosten} by using
an effective action method)%
\begin{eqnarray}
\hbar \omega _{\pm } &=&\frac{1}{2}\left[ \left( 2g-1\right) U+\left(
g-1\right) ^{2}W-\epsilon _{\boldsymbol{k}}-2\mu \right] \pm   \notag \\
&&\frac{1}{2}\left[ U^{2}+\left( g-1\right) ^{2}W^{2}+2\left( g-1\right)
UW-\right.   \notag \\
&&\text{ \ \ }\left. 2\left( 2g+1\right) \epsilon _{\boldsymbol{k}}U-2\left(
2g+1\right) \left( g-1\right) \epsilon _{\boldsymbol{k}}W+\epsilon _{%
\boldsymbol{k}}^{2}\right] ^{1/2},  \notag \\
&&  \label{dispersion relation}
\end{eqnarray}%
where $\epsilon _{\boldsymbol{k}}=2t\sum_{i=1}^{d}\cos \left( k_{i}a\right) $%
, "$\pm $" denote excitations of\ quasi-particles (adding particles to
sites) and quasi-holes (removing particles from sites), respectively.
Furthermore, by subtracting the two solutions, we can obtain the energy gap
\begin{eqnarray}
\Delta _{\boldsymbol{k}}^{2} &=&U^{2}+\left( g-1\right) ^{2}W^{2}+2\left(
g-1\right) UW-  \notag \\
&&2\left( 2g+1\right) \epsilon _{\boldsymbol{k}}U-2\left( 2g+1\right) \left(
g-1\right) \epsilon _{\boldsymbol{k}}W+\epsilon _{\boldsymbol{k}}^{2}.
\notag \\
&&  \label{Ek}
\end{eqnarray}

At the transition point, $\Delta _{\boldsymbol{k}}^{2}=0$ and we have%
\begin{equation}
W_{C}=\frac{-U_{C}+\left[ 2g+2\sqrt{g\left( g+1\right) }+1\right] \epsilon _{%
\boldsymbol{k}}}{g-1}.  \label{Wc}
\end{equation}%
This equation\ shows the $g$-dependence $U_{C}$ and $W_{C}$: as $g$ grows,
the slope $\left( g-1\right) ^{-1}$ decreases; so that $U_{C}$ increases
while $W_{C}$ declining to a fixed value at about $4\epsilon _{\boldsymbol{k}%
}$. For $\boldsymbol{k}=0$, $\epsilon _{\boldsymbol{k}}=zt$, the dispersion
relation(\ref{dispersion relation}) becomes the phase boundary equation(\ref%
{eqn12}), while $\Delta _{\boldsymbol{k}}^{2}$ turning to $z^{2}t^{2}\Delta $%
. Let $\Delta =0$, the relation between $\overline{U}_{C}$ and $\overline{W}%
_{C}$ is%
\begin{equation}
\overline{U}_{C}+\left( g-1\right) \overline{W}_{C}=2g+1+2\sqrt{g(g+1)}.
\label{Uc and Wc}
\end{equation}%
When $g\rightarrow \infty $ and $\overline{U}_{C}=0$, $\overline{W}%
_{C}\approx 4$; when $g=1$, $\overline{W}_{C}$ is divergent and $\overline{U}%
_{C}\approx 5.83$.

\begin{figure}
\begin{center}
\includegraphics[width=0.5\textwidth]{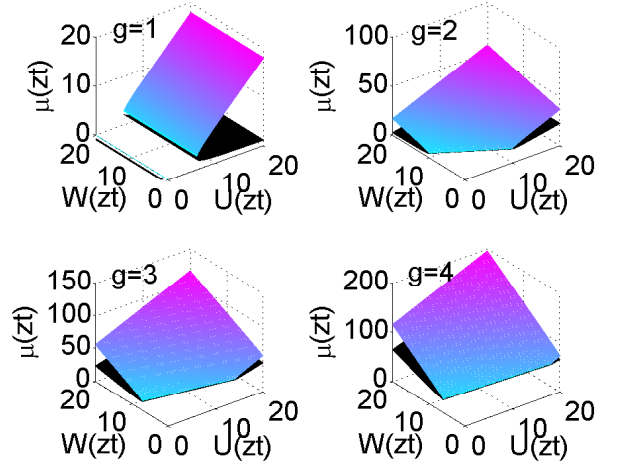}
\caption{(Color online) Mott-insulating 'lobes' in 3D phase space. The areas
enclosed within lobes indicate insulating phase. Space outside these curved
surfaces is superfluid phase.}
\label{Fig.2}
\end{center}
\end{figure}

The physical meanings of this rotation of critical lines may be easier to
comprehend when we plot the complete phase diagrams in three-dimensions in
Fig. \ref{Fig.2}. The well known Mott insulating 'lobes' now expand to a set
of curved surfaces, meanwhile the transition points become the critical
lines in Fig. \ref{Fig.1}. As $g$ increases, these surfaces rotate with
respect to the fixed point $F(5.83,4.01)$.

\begin{figure}
\begin{center}
\includegraphics[width=0.5\textwidth]{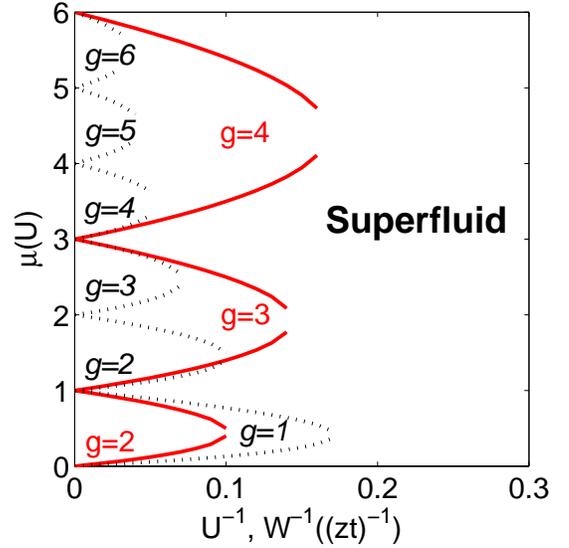}
\caption{(Color online) Phase diagrams of superfluid-Mott insulator
transition with pure two-and three-body interactions. The red lines are
associated with pure three-body interaction and the black dot lines are
associated with pure two-body interaction.}
\label{Fig.4}
\end{center}
\end{figure}

Figure \ref{Fig.4} shows the superfluid phase and Mott insulator lobes of
condensate with pure two-body and three-body interactions, respectively. As
we can see, the\ three-body interaction enlarges the areas of Mott lobes
significantly, which is contrary to pure two-body situation, whose lobes
shrink as $g$ increases: According to Eq. (\ref{eqn12}) and (\ref{delta}),
breadths of\ different\ $U$-lobes are $1$ while $W$-lobes' breadths are
expanding as $(g-1)$ in the unit of $\mu /U$ and $\mu /W$, respectively.

\section{THE EXPERIMENTAL PARAMETERS AND SUGGESTIONS}

The trapping potential is the sum of a homogeneous periodic lattice
potential formed by three orthogonal, independent standing laser fields with
a tunable barrier height $V_{0}$ \cite{W.Zwerger's Review}. For a deep
optical lattice this sine-like trap can be approximated as a harmonic well
as
\begin{eqnarray}
V(x,y,z) &=&V_{0}\left( \sin ^{2}kx+\sin ^{2}ky+\sin ^{2}kz\right)   \notag
\\
&\approx &V_{0}k^{2}\left( x^{2}+y^{2}+z^{2}\right) \text{,}  \label{V0}
\end{eqnarray}%
where $k=2\pi /\lambda $ is the wave vector with $\lambda $ the wave length
of the laser. The Wannier function at the lowest band in the well can be
approximated as a Gaussian ground-state
\begin{equation}
w(\boldsymbol{r})=\left( \frac{\alpha ^{2}}{\pi }\right) ^{\frac{3}{4}}e^{-%
\frac{1}{2}\alpha ^{2}\boldsymbol{r}^{2}}\text{,}  \label{Wannier function}
\end{equation}%
where $\alpha ^{-1}=\left( 2mV_{0}k^{2}/\hbar ^{2}\right) ^{-\frac{1}{4}}$
is the characteristic length of harmonic oscillators. The interaction
parameter $U$ and $W$ are thus given by integrals over the Wannier function
\cite{W.Zwerger}
\begin{equation}
U=g_{2}\int \text{d}\boldsymbol{r}\left\vert w(\boldsymbol{r})\right\vert
^{4}=\sqrt{\frac{8}{\pi }}ka_{s}E_{r}\left( \frac{V_{0}}{E_{r}}\right) ^{%
\frac{3}{4}}\text{,}  \label{U}
\end{equation}%
and
\begin{eqnarray}
W &=&g_{3}\int \text{d}\boldsymbol{r}\left\vert w(\boldsymbol{r})\right\vert
^{6}=g_{3}\left( \frac{V_{0}}{E_{r}}\right) ^{\frac{3}{2}}\left( \frac{k^{2}%
}{\sqrt{3}\pi }\right) ^{3}  \notag \\
&=&\left( \sqrt{3}\pi \right) ^{-3}\ln \left( C\eta ^{2}\right) \frac{%
ma_{s}^{2}}{\hbar ^{2}}U^{2}\text{,}  \label{W}
\end{eqnarray}%
where $g_{2}=4\pi \hbar ^{2}a_{s}/m$ is the coupling constant of the
delta-type repulsion between two\ bosons with $a_{s}$ the s-wave scattering
length and $m$ the mass of bosons, while $E_{r}=\hbar ^{2}k^{2}/2m$ is the
recoil energy. The three body coupling constant has been calculated in a
previous work \cite{T.T.Wu} as
\begin{equation}
g_{3}=\frac{16\pi \hbar ^{2}}{m}a_{s}^{4}\ln \left( C\eta ^{2}\right) \text{,%
}  \label{g3}
\end{equation}%
where $\eta =\sqrt{\left\vert \psi \right\vert ^{2}a_{s}^{3}}$ is the dilute
gas parameter and the constant $C$ in the argument of the logarithm can be
determined by applying a microscopic description demonstrated in \cite{T.
Kohler}. In the limit $V_{0}\gg E_{r}$, the tunneling amplitude $t$ can be
obtained from the exact result for the width of the lowest band in the
one-dimensional Mathieu equation as \cite{W.Zwerger}
\begin{equation}
t=\frac{2}{\sqrt{\pi }}E_{r}\left( \frac{V_{0}}{E_{r}}\right) ^{\frac{3}{4}%
}e^{-2\sqrt{\frac{V_{0}}{E_{r}}}}\text{.}  \label{t}
\end{equation}

Combining Eq. (\ref{U}), (\ref{W}) and (\ref{t}), we obtain the relation
between dimensionless interaction strengths $\overline{W}$ and $\overline{U}$
as%
\begin{equation}
\overline{W}=\left( 3\pi \right) ^{-\frac{3}{2}}\ln \left( C\eta ^{2}\right)
\left( \frac{V_{0}}{E_{r}}\right) ^{\frac{3}{4}}e^{-2\sqrt{\frac{\hat{V}_{0}%
}{E_{r}}}}a_{s}^{2}k^{2}\overline{U}^{2}\text{,}  \label{Wbar}
\end{equation}%
where $a_{s}^{2}k^{2}$ ranges from $10^{-8}$ to $10^{-2}$ in current
experiments \cite{Book}. As we can see, in this context, the strength of
three-body interaction is much weaker than its two-body counterpart, which
means that three-body effects can hardly be observed. This is consistent
with the current experiments where the manipulation of two-body interactions
$\overline{U}$ is accomplished through changing the depth of the optical
well $V_{0}$.

However, as Ref. \cite{H. P. Buchler} pointed out, by loading the polar
molecules into the optical lattice and adding an external microwave field,
one can achieve a situation where only observable three-body interaction
exists. It may open a promising route for an experimental study of
controlling different interatomic repulsions individually, and the system
can be possibly prepared with pure three-body interaction. Moreover the
experimental techniques like Feshbach resonance for molecules \cite{Jin} and
the rapid progress in dipolar condensate \cite{Pfau} provide us more
opportunities to tackle the many-body problems in cold atomic gases. In that
case, the above discussions on the exotic quantum phases and the new type of
Mott-Hubbard transition can be examined.

\section{CONCLUSIONS}

In this paper, the superfluid-insulator transition with many-body
interactions between optically-trapped ultracold bosonic atoms is discussed.
We find the extension of the Mott-insulating areas and the existence of a
fixed point in phase space. Finally, we explore the possibility of realizing
the theoretical predictions and suggest some experimental means to test our
results.

\begin{acknowledgments}
This work is supported by NCET, NSF of China under Grant No.
10574014 and 10774095, SRF for ROCS, SEM, 973 Program under Grant
No. 2006CB921102 and Shanxi Province Youth Science Foundation under
Grant No. 20051001.
\end{acknowledgments}

\end{document}